\documentclass[aps,prb,twocolumn,citeautoscript]{revtex4}
\usepackage{amsmath} \usepackage{epsfig}
\begin{document}
\title{Interpretation of the de Haas - van Alphen experiments in MgB$_{2}$}
\author{I. I. Mazin}
\email{mazin@dave.nrl.navy.mil}
\affiliation{ Code 6391, Naval Research Laboratory, Washington, DC
20375-5000, USA}
\author{Jens Kortus}
\email{j.kortus@fkf.mpg.de}
\affiliation{Max-Planck-Institut f{\"{u}}r Festk{\"{o}}rperforschung,
Heisenbergstr. 1, D-70569 Stuttgart, Germany}
\date{\today}

\begin{abstract}
Recent reports on quantum oscillations in MgB$_{2}$ provide valuable
information on three important aspects of this material: (1) electronic
structure near the Fermi level, (2) disparity of the electron-phonon
interaction between the two systems of bands (3) renormalization of spin
susceptibility. However, extraction of most of this information requires
highly accurate band structure calculations of the relevant quantities. In
this paper we provide such calculations and use them to analyze the
experimental data.
\end{abstract}
\maketitle

MgB$_{2},$ a novel superconductor with $T_{c}\approx 40$ K, has attracted
enormous attention in the last year. The most popular model, suggested by
Liu {\it et al.} \cite{Amy} and Shulga {\it et al.} \cite{shulga},
and elaborated by Choi {\it et al.} \cite{Choi}, is the
 two-gap model, which, based on  the
 very large interband disparity of
the electron-phonon interaction (first noted in Ref.\ 
\onlinecite{PKB}), predicts two different gaps
for the two different band systems. The calculations\cite{Amy,Choi}
 yield an effective
(including an enhancement due to gap variation) electron-phonon coupling
constant of the order of 1. 
On the other hand, the two-gap theory has a serious conceptual problem: 
Two distinctive
gaps may exist only 
if the interband impurity scattering is very weak. 
That seems to be in contrast to the experimental observation that
even poor quality, high-resistivity, samples
have very good superconducting properties. It has been argued\cite{Kuzm,imp}
that the specificity of the electronic and crystal structure of MgB$_{2}$
results in a peculiar relation among the three relevant relaxation rates,
namely that the impurity scattering inside the so-called $\pi $ band is much
stronger than inside the $\sigma $ band, and the latter, in turn, is much
stronger than the interband scattering. However, there has been no direct
experimental confirmation of this claim. 

On the other hand, some authors\cite{Mars} argue 
that the calculated band structure is strongly renormalized by
electron-electron interactions not accounted for in the local density
calculations, so that the plasma frequency is a factor of five smaller than
the calculated one. This would imply an  electron-phonon coupling constant less
than 0.2. There are claims that infrared spectroscopy supports this point of
view\cite{optics1,Tu}, although other researchers in the field\cite{Kuzm}
dispute the interpretation accepted in Refs. \onlinecite{optics1,Tu}. 
In any case,
the fact that all optical experiments till now have been performed on
polycrystalline samples, undermines their value as a decisive test for the
electronic structure calculations. 

The first single crystal angular-resolved photoemission (ARPES) \cite{ARPES}
measurements agree very well with  the calculations
\cite{ARPESc}. However, some calculated bands have not been observed, and,
furthermore, ARPES probes only a very thin surface layer and is 
therefore often not representative of the bulk electronic structure.

Historically, the most reliable probe of the bulk electronic structure has
been the de Haas-van Alphen effect (dHvA). Recent observation of this effect in
MgB$_{2}$ single crystals \cite{dhva} provides key information to assess the
validity of the standard band structure calculation. Given the fact that
most theoretical papers rely on this band structure, the importance of a
proper analysis of these data can hardly be overestimated.
It must be emphasized
that such an analysis requires highly accurate band structure calculations,
{\it i.e.,} the use of
 a much finer k-point mesh in the Brillouin zone and a much more
accurate integration than is customary in other applications of the band
theory. In this paper we present such calculations and show that both
Fermiology and effective masses (and hence the Fermi velocities and plasma
frequencies) produced by conventional band structure calculations are in
excellent agreement with the experiment, thus giving a strong foundation for
the widespread use of this band structure. Furthermore, we show that the
calculational predictions of a strong disparity of the electron-phonon
interaction in the two band systems in MgB$_{2}$ are supported by the de
Haas-van Alphen experiment, and that the scattering rates inside the $\sigma 
$ band and between $\sigma $ and $\pi $ bands are probably
much smaller than inside
the $\pi $ bands.

The Fermi surface of MgB$_{2}$ consists of four sheets\cite{Kortus}. Two
sheets come primarily from the boron $p_{x}$ and $p_{y}$ states, and form
 slightly (nearly sinusoidally) warped cylinders, $\sigma $
(bonding) and $\sigma ^{\ast }$ (antibonding)\cite{sigma},
 and two tubular networks, the
bonding one, $\pi ,$ in the $\Gamma $ ($k_{z}=0)$ plane, and the antibonding
one, $\pi ^{\ast },$ in the A ($k_{z}=\pi /c)$ plane. There are 6 extremal
cross-sections for the field parallel to $k_z$ (along the $\Gamma-$A line).
These are: (1) $\sigma $ in the $%
\Gamma $ plane; (2) $\sigma ^{\ast }$ in the $\Gamma $ plane; (3) $\pi $ in
the $\Gamma $ plane (``holes'' between the tubes); (4) $\sigma $ in the A
plane; (5) $\sigma ^{\ast }$ in the A plane; and (6) $\pi ^{\ast }$
in the $\Gamma $ plane. For a field parallel to $k_y$ (perpendicular
to the $\Gamma-$AM plane) there are two extremal
cross-sections (tubes' necks), for the $\pi $ surface (7) and for the $\pi
^{\ast }$ surface (8). 

We performed highly accurate and well converged 
full potential linear augmented plane wave
(LAPW) calculations, using the WIEN-97 package\cite{WIEN}, 
including local orbitals\cite{Singh} to relax the linearization errors.
We used the Generalized Gradient Approximation of Perdew-Wang \cite{GGA} 
for the exchange-correlation potential. 
By comparing
the results with LMTO (linear muffin tin orbitals) calculations,
we found that
for a proper description of the $\sigma $ orbits 
it is essential
to use a full potential method. 
It is furthermore essential to use a very fine mesh in {\bf k}-space; 
we employed a 38x38x27 mesh, corresponding to 1995 inequivalent {\bf k}-points. 
To achieve sufficient accuracy for
 the small areas of the orbits 1,2,4,5,7 and 8,  
we used an integration engine built 
in the SURFER program \cite{surf}, which internally interpolates the
integrand with splines. 

\begin{table*}
\caption{\label{tab1}
Calculated de Haas-van Alphen parameters from present work 
($F_{\text{calc}}$) 
compared to the experimental data ($F_{\text{exp}}$) of 
Ref. \protect\onlinecite{dhva}. The masses are given in  
free electron mass units.} 
\begin{ruledtabular}
\begin{tabular}{cl|rrcrrrr}
\multicolumn{2}{c}{Orbit} & $F_{\text{calc}}$
 [T] & $m^{calc}$&$dm^{calc}/dE$ [Ry$^{-1}$] 
 & $\lambda$\footnote{Computed from Tables 1 and 2 of Ref. \protect\onlinecite{Amy}.}
 & $|(1+\lambda)m|$\footnote{Computed from the preceding columns}
 &  $F_{\text{exp}}$
  [T] 
 & $|(1+\lambda )m|^{\text{exp}}$ \\ 
\hline
1&$\sigma$ $\Gamma$-plane& 730& -0.251& 1.1&1.25& 0.56 & 
                                               540& 0.54 \\ 
2&$\sigma^*$ $\Gamma$-plane& 1589& -0.543&2.7 & 1.16& 1.17& 
                                                      &  \\ 
3&$\pi$ $\Gamma$-plane & 34630 & 1.96&23
& 0.43 & 2.80 & 
                                                      &  \\ 
4&$\sigma$ A-plane & 1756& -0.312 &1.2& 1.25& 0.70& 
                                              1530 & 0.66 \\ 
5&$\sigma^*$ A-plane& 3393 & -0.618&2.3 & 1.16 & 1.33 & 
                                                       &  \\ 
6&$\pi^*$ $\Gamma$-plane & 31130& -1.00&4.1 & 0.47& 1.47& 
                                                       &  \\ 
7&$\pi$ $\Gamma$AM-plane  & 458 & -0.246 &1.5& 0.43 & 0.35   &  &  \\ 
8&$\pi^*$ $\Gamma$AM-plane  & 2889 & 0.315 &0.8& 0.47 & 0.46 & 2685 & 0.45%
\end{tabular}
\end{ruledtabular}
\end{table*}

The bare (band) masses in the third column of table I were 
then calculated by varying the Fermi energy and using the 
standard formula, 
\begin{equation}
m_{\text{dHvA}}=\frac{\hbar^2}{2\pi} \frac{dA}{dE}.
\end{equation}
Here and below we use the notation $A$ for the 
areas of the orbits in standard units and $F$ for those in Tesla units.
In order to obtain the energy derivatives we fitted
the calculated $A(E)$ by quadratic polynomials in
the ranges of about 0.03 Ry around the Fermi energy.
The experimentally observed ``thermal masses'' 
differ from the ``band'' masses by a renormalization factor of 
$(1+\lambda)$,
where $\lambda$ is the coupling constant for the interaction
of electrons with phonons or other low-energy excitations.
For Table I we used the values of $\lambda$ computed in the following 
way (see, {\it e.g.}, Ref. \onlinecite{phys}): 
we assumed that the matrix elements of the electron-phonon interaction 
are constant within each of the 4 bands (a good approximation, see
Ref.\onlinecite{Choi}), but different among the bands and for 
different interband transitions. 
If the matrix of the electron-phonon interaction is $U_{ij}$, 
where $i,j$ are the band indices, then the mass renormalization in the band 
$i$ is 
\begin{equation}
\lambda_i=\sum_j{U_{ij}N_j},
\end{equation}
where $N_j$ is the partial density of states per spin for 
the $i$-th band. Recall that the conventional Eliashberg coupling 
constant is $\lambda=\sum_{ij}{U_{ij}N_iN_j}/\sum_i{N_i}$.
The matrix $U$ and the vector $N$ calculated in Ref.\ \onlinecite{Amy}
were used to compute the fifth column in Table I.    

The agreement between the calculated and measured thermal masses can be
characterized as excellent. Very importantly, {\it this agreement is so
good only because the calculated electron-phonon coupling differs by a
factor of 3 between the }$\sigma $ {\it and }$\pi $ {\it bands.} This is the
first direct demonstration of this important effect. The agreement between
the calculated areas $F$
and the experiment is also very good. Although $%
F_{1},$ $F_{2}$ and $F_{3}$ are overestimated by 35\%, 15\%, and 8\%,
respectively, 
 the absolute values of these errors
are only 0.5\% (or less) of the total area of the corresponding 
Brillouin zone cross-sections. Even better appreciation of
the significance of these errors can be gained from the 
observation that 
shifting the $%
\sigma $ band by 6.3 mRy down, and the $\pi ^{\ast }$ band by 5.5
 mRy up brings the 
calculated areas to full agreement with the experiment. 
It is not at all clear whether or not such a small discrepancy
with the experiment is meaningful. 
It is interesting, nevertheless, that
after such an adjustment of the band positions the calculated masses
agree with the experiment even better: for the three
orbits in question the
electron-phonon coupling constants deduced from the experiment by taking 
the ratio of the measured masses to the calculated masses are, respectively,
1.15,1.12, and 0.43. After the 
Fermi level adjustment, they  are 1.22, 1.18, and 0.45.
It is also worth noting that, for instance, a change in c/a ratio
of 1.5\% shifts the $\sigma$ and $\pi$ bands with respect to each other
by $\approx 12 mRy$, or that a shift of the Fermi level by 6 mRy corresponds
to a 0.05 $e$ change in the number of electrons. This shows how sensitive
the de Haas-van Alphen results are to the crystallography and stoichiometry. 

Another important observation reported in Ref.\ \onlinecite{dhva} is 
the so-called ``spin-zero''. This is a suppression of the de Haas-van Alphen 
amplitude when the difference in the areas (in Tesla units) of the spin-split 
(by the external field $H$) cross-sections is exactly $H/2$. 
This effect has been observed for orbit 8 in the field $H=17$ T, 
when the field was tilted with respect to the crystallographic 
axis by $\phi =15-18^{\circ }$. This
means that ($F_{8}^{\uparrow }-$ $F_{8}^{\downarrow })/\cos (\phi )=8.5$ T,
or  $\Delta F_{8}=F_{8}^{\uparrow }-$ $F_{8}^{\downarrow }\approx 8.1$ T
(note that the angle itself does not depend on the field in which 
the measurements are performed, but only on the Fermi surface geometry 
and Stoner renormalization). It
is easy to estimate this splitting in the first approximation, using the data
from the Table I and the Stoner renormalization of 33\%, calculated in Ref.\
\onlinecite{eva}: $\Delta A_{8}=2\pi m\Delta E_{xc},$ where 
$\Delta E_{xc}=2\mu _{B}H(1+S)$ is the induced spin-splitting of the 
bands near the Fermi level, enhanced by a Stoner factor $(1+S)$. 
This formula gives $\Delta F_{8}\approx 7.1$ T. 
A { caveat} here is that the induced spin-splitting need not be the same
for all bands, in other words, while the {\it average }$S$ is 0.33,
individual $S$'s may vary from orbit to orbit. To avoid this problem, we
performed self-consistent LAPW calculations in an external field of 1.8 kT
(still well within the linear response regime) and measured 
$\text{d}\Delta A_{8}/\text{d}H$
explicitly. Using these results, we found that for the actual field of 17 T
$\Delta F_{8}= 6.7$ T, close to, but slightly smaller than the above
estimate of 7.1 T. In other words, the calculated Stoner factor for this orbit
is $S_8=0.26$, smaller than the average over all bands, which is 0.33. Note that
the experimental number of 8.1 T can be reconciled with the calculated mass,
if $S_8$ were $\approx 0.5,$ fairly close to the electron-phonon coupling 
constant for the same band, 0.47. 
We,
however, believe that the coincidence is accidental, although we do
not have any plausible explanation for the noticeable underestimation
of the Stoner factor for this orbit.
No ``spin-zero'' effect
has been observed for the orbit 4, which has essentially
the same mass as orbit 8. Our calculations for this orbit give $\Delta 
F_{4}= 6.9$ T; 
that is, the calculated Stoner factor for this orbit is $S_4=0.31$.
 At the same time, the actual Stoner factor
must be either larger than 0.60 or smaller than 0.18, for this orbit not to 
exhibit the ``spin-zero'' effect (this is neglecting  deviations
from a cylindrical shape, which are noticeably stronger
expressed for this orbit than for the orbit 8). 
Further experimental studies on better
samples should give more insight into this problem.

Finally, we would like to discuss the problem of the ``missing orbits''. The
amplitude of the de Haas-van Alphen signal is proportional to\cite{wasser}:  
\begin{eqnarray*}
&&H^{-\frac{1}{2}}\frac{X}{\sinh X}\exp \frac{-c\hbar \sqrt{\pi A}}{%
eH\ell }\cos \frac{\pi \Delta F}{H} \\
X &=&\pi ^{2}mc(1+\lambda )k_{B}T/\hbar e H,
\end{eqnarray*}%
where $\ell $ is the mean free path for the orbit in question. Thus, it is
not surprising that the large orbits 3 and 6 are not observed; the Dingle
exponent $c\hbar \sqrt{\pi A}/eH \ell $ is at least 10 times larger
than for the other orbits. However, the question remains for the orbits 2, 5,
and 7. Let us start with the first two. We observe that, compared to the
orbits 1 and 3,  both Dingle factor and the thermal factor are reduced. The
latter is smaller because the effective mass, $m(1+\lambda )$ is twice
larger, which reduces the maximal temperature at which these orbits can be
observed by a factor of two. The former is reduced because both the orbit
size, $\sqrt{A}$, is larger, and the mean free path, $\ell \propto v_{F},$
(assuming the relaxation time is the same for both $\sigma$ and $\sigma^*$
bands)
is smaller (from Table 1 of Ref. \onlinecite{eva}, $v_{F}(\sigma )/v_{F}(\sigma
^{\ast })\approx 1.4$). The total reduction of the Dingle exponent compared
to orbit 4 is by a factor of 2 for orbit 5, and of 1.4 for orbit 2.  

The absence of a signal from the orbit 7 seems puzzling. Its area and its
thermal mass are the smallest of all orbits, and the average velocity for
this band is the highest (50\% higher than for the $\sigma $ band). 
A very plausible explanation is that, 
as conjectured in Ref.\ \onlinecite{Kuzm} and elaborated 
in Ref.\ \onlinecite{imp}, the impurity 
scattering rates differ drastically between the bands. 
If the dominant 
defects reside in the Mg plane ({\it e.g.}, Mg vacancies), 
then
such defects are very weak scatterers for the $\sigma $ bands for 
the simple reason that those bands have very little weight at the Mg atoms. 
However, this simple picture does not explain
why orbit 8, originating from the  $\pi ^{\ast }$ band, apparently has
a small relaxation time and therefore is seen in experiment. 
Its velocity is close to (in fact, 15\% smaller than)
that of the $\pi $ band, its linear size is more than twice larger than that
of orbit 7, so the scattering rate has to be at least 5 times larger. 
We do not have a plausible answer at the moment why the
impurity scattering appears to be so suppressed for this orbit. Possibly,
this is related to its parity (while the $\pi $ band is even with respect to
the $z\rightarrow -z$ reflection, the $\pi ^{\ast }$ band is odd).

To conclude, we presented highly accurate calculations of the de Haas-van
Alphen parameters for MgB$_{2}.$ Comparison with the experiment reveals: (a)
Absence of any mass (velocity) renormalization apart from that due to
phonons. (b) A good agreement of the calculated cross-section areas with the
experiment. 
(c)
Excellent agreement of the calculated electron-phonon coupling with the dHvA
mass renormalization, including very large disparity between the coupling of
the $\sigma -$ and $\pi -$bands, which 
clearly confirms the basic assumption of 
the two-gap model for superconductivity in MgB$_2$.
(d) Some underestimation, despite a good
qualitative agreement, of the calculated and measured Stoner factors for the 
$\pi $ bands. (e) An indirect evidence of substantially different impurity
scattering rates in the $\sigma $ and $\pi $ bands. (f) A problem which
remains to be understood is the total suppression 
of the neck orbit, associated with the bonding $\pi $ band, 
given a clear observation of the much larger orbit from the 
electronically similar $\pi ^{\ast }$ band.

After this work was finished,
we learned about similar works 
by Rosner {\it et al.}
\cite{Rosner} and Harima \cite{Harima}. Their results, particularly
those of Ref. \onlinecite{Rosner}, are quite close to ours.
Both paper employ similar methods and take full care  of
the k-mesh convergence. The remaining difference is
a good gauge of how reliable are such calculations, in
technical sense.

We are grateful for A. Carrington and J. R. Cooper for numerous extremely
enlightening discussions regarding their paper\cite{dhva}, as well as to 
O. K.  Andersen, O. Jepsen and O. V. Dolgov for many discussions of 
the electronic structure and transport properties of MgB$_{2}$. 
We also thank R. Hayn, H. Rosner, and S.-L.  Drechsler for their useful comments.

JK would like to thank the Schloe{\ss}mann Foundation for financial support. 
The work was partially supported by the Office of Naval Research.


\begin{thebibliography}{99}
\bibitem{sigma} Formally speaking, there are four $\sigma$ bands, of
which two are bonding in the sense of covalent bonds, and two antibonding.
Of the two bonding bands, one is more bonding than the other (except at
the $\Gamma$ point) due to
a favorable combination of $p_x$ and $p_y$ orbitals. In this paper, we call 
this band $\sigma$, and the other $p_{x,y}$ band, crossing the Fermi level,
$\sigma^*$.
\end{thebibliography}
\end{document}